# PLANAR VELOCITY ANALYSIS OF DIESEL SPRAY *SHADOW IMAGES*

D. Sedarsky[§], S. Idlahcen, J-B. Blaisot, C. Rozé
CORIA, Rouen University, Rouen, France
[§]Correspondence author. Email: david:sedarsky@coria.fr

**ABSTRACT**   The focus of this work is to demonstrate how spatially resolved image information from diesel fuel injection events can be obtained using a forward-scatter imaging geometry, and used to calculate the velocities of liquid structures on the periphery of the spray.

Useful visual information on the structure of diesel sprays is often obtained by projecting intense source light across the volume of interest such that the transmitted intensity forms a line-of-sight integrated measurement. The forward-scattered intensity collected from such an arrangement is commonly known as a shadowgram, and contains structural information from the measured volume in the form of dark regions imprinted on the source light background level.

 Much useful information can be gained by observing a diesel spray in this configuration, however, the resulting signal does not form an image in the conventional sense. The collected 2-D spatial information is a projection of the shadows formed by refractive index gradients in the measurement volume. As such, the character of the phase information at the detection plane is not well-ordered and cannot be interpreted or manipulated exactly in the manner appropriate for true images. In addition, the shadowgram spatial information is line-of-sight integrated, providing no appreciable spatial resolution along the optical axis of the measurement.

 In order to obtain accurate velocities directly from individual diesel spray structures, those features need to be spatially resolved in the measurement. The distributed structures measured in a direct shadowgraphy arrangement cannot be reliably analyzed for this kind of velocity information. However, by utilizing an intense collimated light source and adding imaging optics which modify the signal collection, spatially resolved optical information can be retrieved from spray edge regions within a chosen object plane.

 This work discusses a set of measurements where a diesel spray is illuminated in rapid succession by two ultrafast laser pulses generated by a mode-locked Ti-Sapphire oscillator seeding a matched pair of regenerative amplifiers. Light from the object plane, located at the center of the spray, is directed by a set of imaging optics to the detection plane, where it forms a *shadow image* on the face of a frame-transfer CCD. In this configuration the depth-of-field of the imaging optics limit detected light, effectively increasing the efficiency of signal collection within the object plane and suppressing out-of-focus contributions. Time-correlated images are collected in this manner and analyzed to obtain object plane velocity data for in-focus edge features of the spray.

*Keywords* – imaging, diesel spray, velocity, spatially resolved, shadowgram

## NOMENCLATURE

| | |
|---|---|
| x — optical axis, distance coordinate | h — shadowgram to light source distance |
| y — vertical axis, distance coordinate | g — shadowgram to object plane distance |
| z — horizontal axis, distance coordinate | n — spatially varying refractive index |
| ε — refraction angle | $E_0$ — intensity |
| ΔE — intensity change | k — wave vector |
| a — aperture distance | r — image plane radius |
| f — focal length | δ — defocus distance |
| U — far-field amplitude | λ — wavelength |
| $J_1$ — Bessel function | I — far-field intensity |
| I — corr. search region | T — corr. template region |
| σ — standard deviation | n — no. image pixels |
| $s_i$ — image distance | $s_o$ — object distance |
| $X_N$ — near image distance | $X_F$ — far image distance |
| $D_N$ — near image DoF distance | $D_F$ — far image DoF distance |
| $X_N$ — near image distance | $X_F$ — far image distance |
| $\Phi_{circ}$ — circle of least confusion | |

## INTRODUCTION

Diesel combustion devices are essential components of the modern world which are widely used in transportation, power generation, construction, and many other industries. This widespread use and the efficiency advantages of diesel combustion provide a large incentive for realizing improvements in diesel combustion designs. The combustion process, and hence, diesel engine performance is largely driven by the character of the fuel/air mixture generated in each cycle by the high-pressure spray used disperse the liquid fuel. This injection process prepares the reactant charge, allowing combustion to occur efficiently in the vapor phase. Important aspects of this mixing process remain unclear due to difficulties in resolving the temporal and spatial scales relevant to the high-pressure, high-speed, multiphase fluid dynamics of a diesel spray in the midst of a turbulent flowfield. The optically dense nature of the diesel spray frustrates many measurement approaches due to the significant light attenuation and multiple-scattering noise which often interfere with the signal of interest. In this work, we examine the use of a transillumination imaging arrangement, modified to acquire spatially resolved dynamic information from a Common Rail (CR) diesel test injector.

## PROJECTED SHADOWS

Sprays are often evaluated by observing the light intensity which is transmitted through an illuminated spray volume. Shadowgraphy, the simplest of such arrangements, is routinely applied to observe multiphase flow structure. Here, the spray is illuminated and the shadows formed by refractive index variations within the volume are projected on a detection plane positioned opposite the light source. The resulting 2-D light intensity sampled at the detector is termed a 'shadowgram,' and contains a projection of the measurement volume's spatial information, line-of-sight integrated along the optical axis.

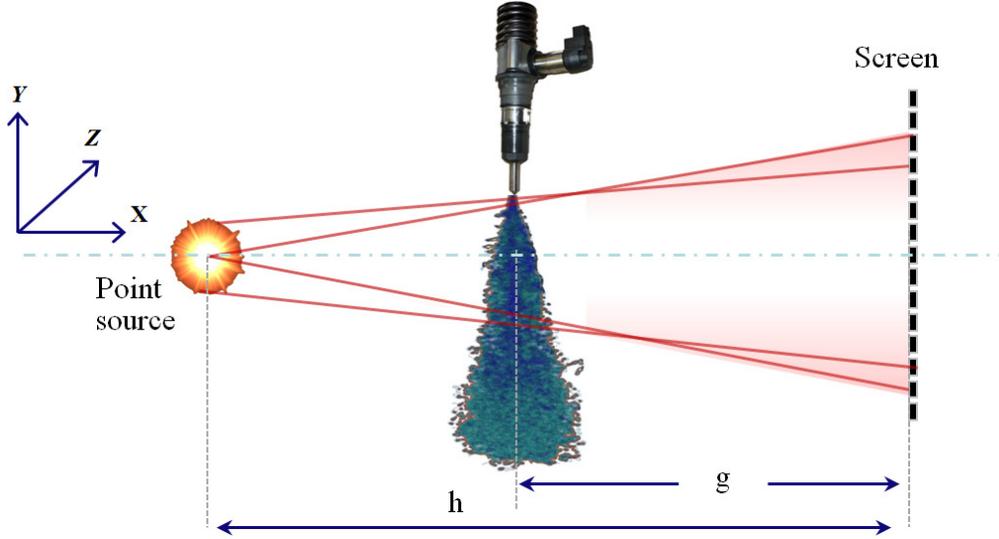

Figure 1. Direct shadowgraphy arrangement

By carefully controlling the properties of the source illumination and collection optics, the projected intensity appears at a known magnification relative to the spray, revealing optical inhomogeneities which are indicative of distributed spray structures. The shadowgram results from the cumulative ray displacements (ray curvature) across the spray region, given by:

$$\frac{\partial^2 y}{\partial x^2} = \frac{1}{n}\frac{\partial n}{\partial y}, \qquad \frac{\partial^2 z}{\partial x^2} = \frac{1}{n}\frac{\partial n}{\partial z} \qquad (1)$$

where $n$ is the spatially varying refractive index. For a direct shadowgraphy system, as shown in Fig.1, the sensitivity of the ray displacement to the refractive index change is given by:

$$\frac{\Delta E_y}{E_0} = \left(\frac{\partial \varepsilon_y}{\partial y}\right)\frac{g(h-g)}{h}, \qquad \frac{\Delta E_z}{E_0} = \left(\frac{\partial \varepsilon_z}{\partial z}\right)\frac{g(h-g)}{h} \qquad (2)$$

where $\varepsilon$ is the refraction angle, and $g$ and $h$ are the distances from the shadowgram to the object and light source, respectively.

The sensitivity of shadowgraphy can be varied by changing the distance, $g$, of the projection to the detection plane (see Eqn.2), allowing the observation of small pressure, temperature, or material variations which produce refractive index gradients. For this reason the technique is widely applied to observe shock waves and temperature variations in transparent media [Izarra 2011, Lewis 1987, Settles 2001].

However, the projected nature of the spatial intensity signal results in a shadowgram, which differs from a true image. Simply stated, a projected shadow does not contain complete information for three-dimensional structure, and structures across the entire line-of-sight contribute indiscriminately to the signal. Although a carefully arranged shadowgram can appear at a quantified magnification with respect to the actual size of the features it measures, the phase information of the detected light is not ordered in a useful sense as it would be in an image. As a result, shadowgrams require some care to handle and interpret properly. For example, the darker regions appearing in a shadowgram, at a gray-level below the steady background level, represent spatial information from the shadowgram scene. However, the bright regions appearing above the background gray-level should not be directly interpreted to represent structure in the scene; in general, these bright regions are formed by caustics (refracted light concentrated in the scene) or noise from stray light. In addition, these difficulties may be further compounded by the spatially-integrated nature of the shadowgram signal which can produce identical shadows for structures positioned at different locations along the optical axis of the measurement.

## IMAGED SHADOWS

By modifying the measurement of the transmitted intensity and taking advantage of the general form of a diesel fuel spray, it is possible to obtain spatially resolved information despite the line-of-sight arrangement of the imaging apparatus.

In general, additional information is required to fully locate 3-D information in a 2-D scene. In planar imaging approaches, this is accomplished by forming a scene where signal is generated only from within the region covered by a thin laser light sheet. For transillumination measurements, this approach is not easily accessible. However, by illuminating the measurement volume with collimated source light and collecting the transmitted signal with a set of optics which gathers light effectively near a chosen object plane and forms an image dominated by object plane light, it is possible to spatially resolve certain types of structure. In this case, resolved spatial information appears as well-focused features in the image, which can be heuristically separated from poorly resolved 'out-of-plane' contributions in the resulting scene.

**2-D image features** High-pressure diesel sprays exhibit an approximate cylindrical symmetry centered on the nozzle orifice. As a consequence, the edges of the spray which intersect an object plane extending from the center of the nozzle are generally not obscured by liquid structures positioned outside of this object plane. In other words, these unobscured edges (including thin ligaments and droplets intersecting the object plane) represent quasi- two-dimensional structures. These structures can be exploited to spatially resolve portions of the three-dimensional volume from the scene generated by light which crosses the spray.

**Imaging optics** By utilizing an intense collimated light source to illuminate the spray and supplementing the direct shadowgraphy arrangement shown in Fig.1 with light collection optics, edge features of the diesel spray which fall within the depth-of-field (DoF) of the imaging system can be resolved and separated from unfocused image regions. These data represent optical information which is spatially resolved along the optical axis of the measurement which can be used to observe and track the velocity and morphology of select structures within the spray volume. Thus, the signal generation advantages of a transmitted light arrangement generating a 2-D spatial intensity can be applied to interpret motion and inform our understanding of the dynamics in the turbulent three-dimensional flow-field of the spray as it develops.

Figure 2 shows such a modified transillumination imaging arrangement. Here, the spray is illuminated with collimated source light and signal from the object plane M is imaged to the collection plane M*. In this configuration the DoF of the imaging optics limit detected light, increasing the relative efficiency of signal collection within the object plane and suppressing out-of-focus contributions [Hecht 1998, Settles 2001].

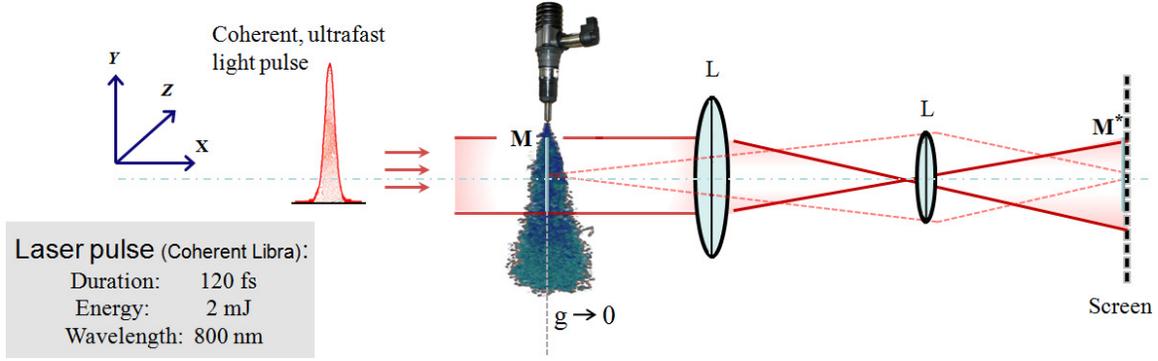

Figure 2. Focused shadow imaging arrangement. Collimated light is directed to the spray. Imaging lenses, L, are positioned to collect light from the object plane, M, at the center of the spray, and image the optical signal to the detection plane, M*.

**Collection optics filtering** The effect of the collection optics can be estimated simply by considering the imaging system in terms of an equivalent thick lens, with a given clear aperture, $a$ (see Fig.3). If the intensity from a distant point source uniformly illuminates the circular lens aperture and a detector is placed a distance $f + \delta$ behind the lens, then the spatial intensity in the detection plane will be given by a circle with radius, $r$. From the similar triangles in Fig.3, one can see that $r$ increases in proportion to the amount of defocusing:

$$\frac{a}{f} = \frac{2 \cdot r}{\delta} \tag{3}$$

It is important to note, since the object and the detector are conjugate imaging planes of the lens system, conclusions drawn from observing defocusing of the detector are equally valid for considering the contributions from displaced object planes to a system with a well-aligned detector.

Disregarding coherence effects, the intensity from an object imaged to the detection plane of the system can be expressed as a spatial convolution of the object with the point-spread-function (PSF, i.e. the geometric impulse response) of the lens, which is equivalent to a Fourier transform (FT) of the lens aperture. For a circular aperture, this FT can be written as a Fourier-Bessel transform [Goodman 2005], yielding the far-field diffraction amplitude,

$$U(r) = \frac{e^{jkx}}{j\lambda x} \exp\left(j\frac{kr^2}{2x}\right)\left(A\frac{J_1(2\pi a \cdot \rho)}{\pi a \cdot \rho}\right) \tag{4}$$

where $\rho$ represents the radial spatial coordinate in the frequency domain,

$$\rho = \sqrt{f_Y^2 + f_Z^2} \tag{5}$$

$A$ is the area of the aperture, and $J_1$ is a first order Bessel function of the first kind. The resulting intensity distribution,

$$I(r) = \left(\frac{A}{\lambda x}\right)^2 \left(2\frac{J_1(k \cdot ar/x)}{k \cdot ar/x}\right)^2 \tag{6}$$

is commonly known as the Airy pattern, and consists of a central peak surrounded by concentric rings of decreasing magnitude. The shape of the Airy pattern ensures that most of the power is concentrated on the central peak, with significantly diminished amplitudes toward the outer, high-frequency regions. Hence, the dominant effect of the transform is to act as a low-pass filter. As the amount of defocusing is increased (increasing $\delta$), $r$ increases, spreading the light from the aperture over a larger area and decreasing the local irradiance at the detector. The first zero of the Bessel function, $J_1(k \cdot ar/x)$ appears closer to the origin, concentrating a larger fraction of the signal power at low frequencies; the attenuation of high spatial frequency components is increased, blurring the detailed structure in the scene. The important point here is that since the main effect of a poor focus is the reduction of high spatial frequency information, the detection of regions with high frequency content is a reliable indicator of well-focused image regions.

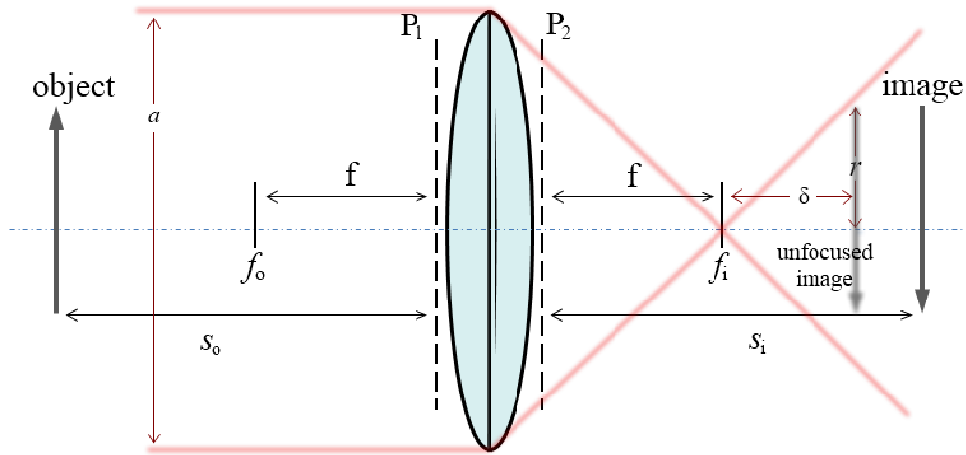

Figure 3. Lens representing arbitrary imaging optics, with focus, f, and principal planes, $P_1$ and $P_2$. The broad red lines represent rays from a distant object which would form a focused image at distance, $f$, or a blurred intensity if the detection plane is positioned at some distance, $f+\delta$, behind principal plane $P_2$.

**In focus regions** The determination of in-focus image regions based on the spatial frequency content of image sub-regions could be accomplished directly by computing the FFT of the image to obtain the spatial frequency distribution [Ng 2001]. Since sharply defined edges contain high spatial frequency information, these regions can be identified by examining the high frequency energy in the power spectrum. However, this represents an appreciable computational burden for large images due to the complexity of the FFT and is not strictly necessary. In practice, it is much more efficient, and quite effective to apply edge detection directly to the image data.

**Depth of field** The relevant DoF pertaining to the measurement volume can be estimated by considering an equivalent lens for the light collection optics, as discussed above. Figure 4 illustrates the DoF, circle of least confusion, and relevant parameters for a lens collecting light from an object at a given distance, $s_o$. Here, the object plane is well-focused at the image distance, $s_i$. Light arriving from in front or behind the object plane at distances $D_N$ and $D_F$ are focused to image distances, $X_N$ and $X_F$, respectively. The DoF is defined by the light-gathering ability of the optics, in relation to the accepted resolution for the imaging system.

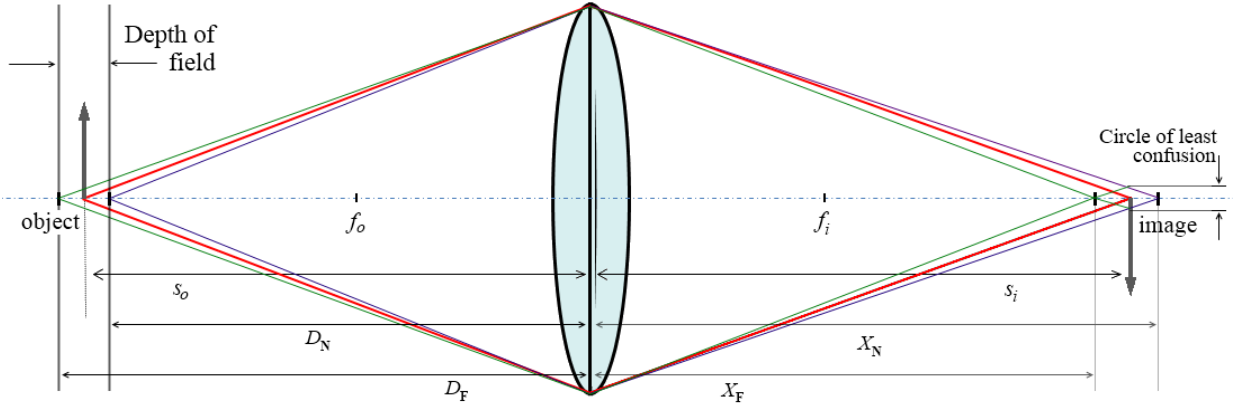

Figure 4. Depth of field (DoF) and circle of least confusion (CLC) for a lens representing light collection optics in a transillumination imaging system. The red lines represent well-focused rays from an object which are correctly directed to the detection plane at distance, $s_i$. The violet and green lines represent rays from the near and far limits of the DoF which are focused to distances $X_N$ and $X_F$, respectively.

Recalling the Gaussian lens law and examining the similar triangles in Fig. 4, one can write

$$\frac{1}{s_o} + \frac{1}{s_i} = \frac{1}{f} \tag{7}$$

$$\frac{X_N - s_i}{X_N} = \frac{X_F - s_i}{X_F} = \frac{\Phi_{circ}}{a} \tag{8}$$

where $\Phi_{circ}$ is the circle of least confusion, $a$ is the clear aperture and $f$ is the focal length of the lens in Fig.4. The equations listed in Eqn. 7 and 8 can be solved for the image distances, $s_i$, $X_N$, and $X_F$, allowing the near and far limits of the DoF to be expressed as

$$D_N = \frac{s_i f^2}{f^2 + \left(\frac{f}{a} \cdot \Phi_{circ} \cdot (s_i - f)\right)} \tag{9}$$

$$D_F = \frac{s_i f^2}{f^2 - \left(\frac{f}{a} \cdot \Phi_{circ} \cdot (s_i - f)\right)} \tag{10}$$

Figure 5 shows the variation of light collection DoF over a range of focusing distances for the transillumination imaging system described in the following section (see collection geometry illustrated in Fig.2).

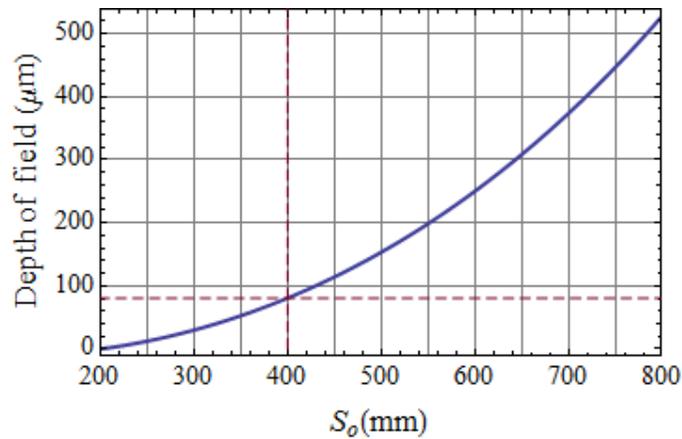

Figure 5. Depth of field (DoF) vs. the focused distance, $s_i$, for the light collection optics shown in Fig.2. Focal length, $f = 200$ mm, clear aperture, $a = 50$ mm. The dashed lines indicate the image distance for unit magnification.

## MEASUREMENT SETUP

Shadow images of a diesel test injector were recorded using an optical arrangement similar to the system described in Fig. 2., utilizing a dual-pulse laser source and a frame-transfer CCD system to allow the acquisition of spatially resolved, time-correlated image data. It is worthwhile to note that the sensitivity of the imaging system differs from that of the direct shadowgraphy system described by Eqn. 2, since the parameter, g, controlling the geometric spread of the ray-curvature signal is approximately equal to zero. As a consequence, significant signal contributions are made only by very strong refractive index gradients. For the work discussed here, the interest is in spray formation and the details of the injection stream dynamics as the emerging fluid shears and breaks into smaller liquid structures. Since all of these structures present very steep refractive index gradients, there is no need for sensitivity to small gradient variations, and this low sensitivity system is well-suited to the present imaging task.

A pair of intense, coherent light pulses from a dual-amplifier laser system were used to illuminate the near field of the diesel injector. The transmitted light was then collected by an objective lens (Nikon, f = 200 mm) and imaged to the detector at approximately 1:1 magnification. The laser source light was generated by a specialized system consisting of two matched regenerative amplifiers seeded by a single Ti:Sapphire laser oscillator. This system is capable of illuminating the spray with a pair of synchronized ultrashort (~120 fs) laser pulses with precisely controlled inter-pulse timing, at a repetition rate of 1 kHz. The intense peak energy and precise timing of the source laser pulses ensure that the system can make full use of dynamic range of the detection optics. In addition, the ultrashort source light serves to precisely freeze the spray motion as well as the optical seeing conditions in the imaging path. This results in sharp spray edges and well-resolved features in regions where the object plane, positioned at the center of the spray, intersects the periphery of the liquid column.

The injector used in this work was mounted on a three-way translation stage housed in an enclosure with four windows, providing line-of-sight optical access in two directions. The measurements were carried out using a calibration liquid (ISO 4113) with properties similar to diesel fuel with carefully controlled viscosity, density, and surface tension specifications (see Table 1). The single-orifice injector with a hole diameter of 113 µm was supplied with fuel through a Common Rail accumulator, allowing adjustable injection pressures from 40 to

100 MPa. Each injection event was controlled electronically by two current levels of 400 µs duration, at a repetition rate of 1 Hz, where the reference clock for the complete system was sourced from the laser oscillator operating at 80 MHz.

## VELOCITY FROM CORRELATED IMAGE-PAIRS

The focus of the work presented here is to measure the velocities of liquid structures on the periphery of a diesel fuel injection spray by comparing time-correlated measurements of the spray.

In order to produce meaningful velocity information, the matched structure in each image-pair must be spatially resolved, i.e. reliably confined to the object plane, or otherwise understood, such that the intensity change in the images can be related to the real-world motion or morphology of the observed structure. Referencing the discussion in the preceding sections, it is clear that properly arranged light collection optics in conjunction with certain features of spray geometries can be exploited to acquire spatially resolved image data which is amenable to analysis for dynamic information.

A variety of methods can be applied to calculate the motion of resolved structure, and hence velocity, from successive images. The current work is based on region-matching method which employs the normalized image cross-correlation given by

$$\frac{1}{n-1} \sum_{y,z} \frac{1}{\sigma_I \sigma_T} \left( I(y,z) - \bar{I} \right) \cdot \left( T(y,z) - \bar{T} \right) \tag{7}$$

Here, 'search field' and 'template' image sub-regions, $I(y,z)$ and $T(y,z)$, are chosen from images produced by consecutive laser pulses which are spaced appropriately in time to produce time-correlated image data.

Similar correlation matching approaches are applied to excellent effect in particle image velocimetry (PIV) [Adrian 2010] and laser speckle imaging (LSI) techniques [Zakhorav 2009]. This approach is limited to some degree by the window sizes which can be practically employed to match any given spatial intensity region, as regions with low entropy or feature content require larger matching regions, or in some cases cannot produce conclusive matching [Kadir 2001]. Nevertheless, the method has the advantage of directly matching visible, human-recognizable fluid structures, facilitating intuitive understanding of the matching algorithm and interpretation of results. In this work it has proven to be effective when applied to the distinct edge features resolved in the ultrafast shadow images.

In addition to the valuable resolved features of droplets and the spray periphery, the shadow images acquired by the optics contain unfocused structure and regions which do not provide reliable data for correlation analysis. Focused shadow image features must be identified and separated from this unfocused background. This is accomplished in two stages, both of which identify high frequency information content in the shadow image spatial intensity allowing focused image regions to contribute to the region-matching analysis and discriminating against regions with poor focus metrics.

Prior to analysis, the images are normalized to the measured image background. The image from the first pulse is taken at time $t_1$; the second pulse produces an image at time, $t_2 = t_1 + \Delta t$, where Δt is given by the time delay between each pair of laser pulses. A set of rectangular image regions are selected from the $t_1$ image, and cross-correlated with a set of rectangular regions selected from the $t_2$ image, to yield the *dy/dt* and *dz/dt* components representing velocity of resolved features confined to the object plane.

The selection of appropriately located $t_1$ image region windows is done first by applying a Sobel edge detection scheme to the shadow image intensity data, identifying candidate regions which should be allowed to contribute to the dynamic analysis. Edge pixels selected by the detection scheme are randomly sampled to form template and search field image sub-regions which are processed to obtain correlation matchings.

In the second stage, the matching results and their associated image sub-regions are validated by a set of criteria which examine the spatial variance and texture energy of the image regions, and the correlation strength and bounds constraints of the matching results. With the exception of the bounds constraints, each of these validation criteria amount to a test of spatial information content for the tested region, with an associated threshold level which further discriminates image regions selected to participate in the analysis.

## VELOCITY RESULTS

To demonstrate the efficacy of the discrimination of focused image regions and subsequent velocity analysis, a pair of time-correlated shadow images of the diesel spray operating at an injection pressure of 40 MPa were sampled at 500 locations and analyzed with region-matching analysis to obtain velocity vectors. The results shown below in Figs. 6 and 7, were calculated for the same set of time-correlated images, with identical template and search field window sizes of 40 x 40 and 120 x 120 pixels, respectively. The image dimensions were 2048 x 2048, at a scale of 0.9 µm/pixel.

For the results shown in Fig. 6, the first focused-region metric, Sobel edge detection, was not used to select the sampled regions. Instead, target pixels were selected on a rectangular grid. As a result, the efficiency of the analysis effort is reduced, since unfocused regions are more abundant than focused regions in the image. The left-hand side of Fig.6 shows unvalidated velocity results, while the right-side shows results validated with the regimen of tests that comprise the second stage of focused-region identification. It is apparent here that the naïve application of the velocity analysis to unfocused image regions, shown on the left side of Fig.6, produces noisy and unreliable velocity information. The normalized cross-correlation method is, to some degree, inherently sensitive to high frequency information [Olsen 2000]. However, this dependence is not strong enough in this case to provide accurate results for diesel spray shadow images without additional steps to eliminate interference from unfocused regions.

The results shown in Fig.7 were obtained using Sobel edge detection to select the 500 sample regions used in the analysis. As before, the left-hand image shows unvalidated and the right shows validated results. It is readily apparent that the selection of focused-regions has improved the results, relative to the vectors shown in Fig.6, even in the unvalidated case (Fig.7, left side). Comparing the left and right-side results of Fig.7, one can see that a number of unfocused regions lead to erroneous vectors in the unvalidated case. This is a consequence of the implementation of the edge detection in a targeting scheme, which has the advantage of improving the efficiency of the analysis effort, but does not actively exclude contributions from unfocused elements.

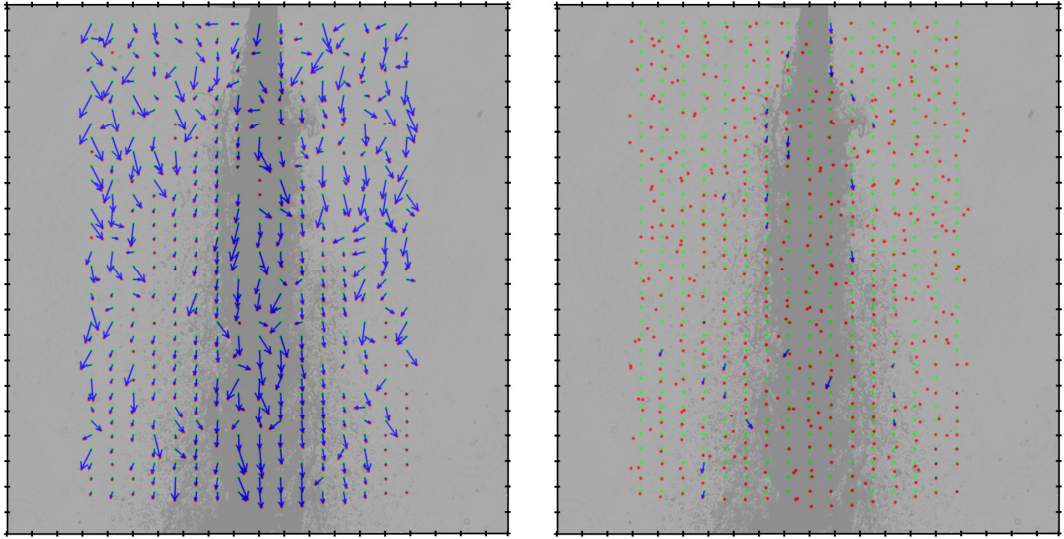

Figure 6. Velocity vectors calculated from time-correlated shadow images of a diesel spray at 500 regularly spaced sample points. The green points mark start coordinates, red points mark the correlation matched end coordinates; velocity vectors are shown as blue arrows. The panel on the left shows results without validation. The panel on the right shows the result when only validated regions with high-frequency spatial information are allowed to form velocity vectors.

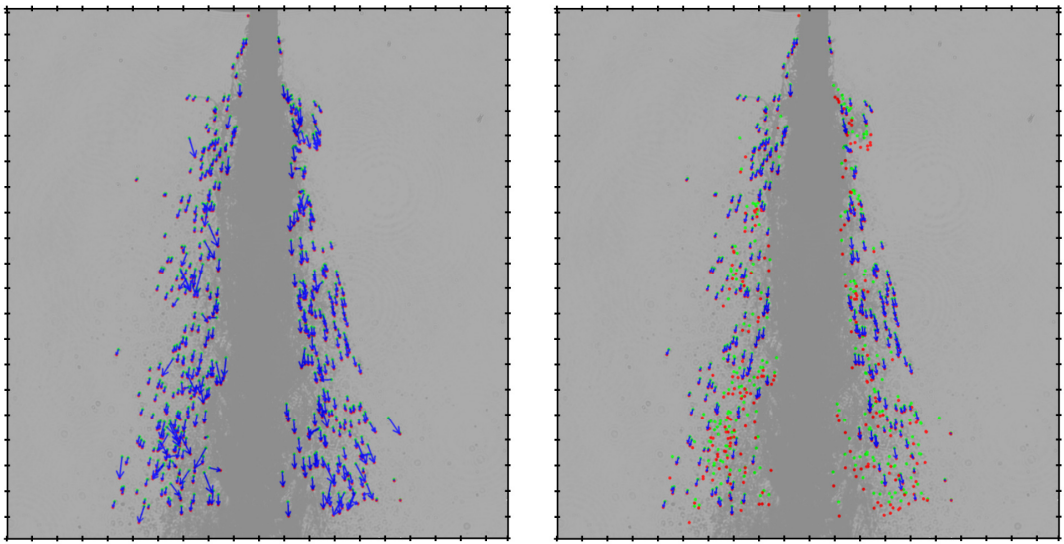

Figure 7. Velocity vectors calculated from time-correlated shadow images of a diesel spray at 500 sample points chosen randomly from pixels identified with a Sobel edge detection method. The green points mark start coordinates, red points mark the correlation matched end coordinates; velocity vectors are shown as blue arrows. The panel on the left shows results without validation. The panel on the right shows the result when only validated regions with high-frequency spatial information are allowed to form velocity vectors.

# CONCLUSIONS

The analysis and experimental efforts presented here demonstrate how spatially resolved image information can be obtained from the optically challenging environment of a diesel fuel injection spray using a forward-scatter imaging geometry adapted from shadowgraphy and planar imaging methods.

By taking advantage of symmetry and spray characteristics, it is possible to obtain spatial data which can be analyzed to quantitatively inform our understanding of spray dynamics. The methods discussed here were successfully applied to calculate planar velocity components for liquid structures on the periphery of a single-hole CR diesel spray, confined to the DoF of the imaging system.

These results show that accurate velocities can be obtained in regions near the periphery of high-pressure diesel injection events, provided that the spray exhibits significant resolved features which can be discriminated from the unfocused background noise.

Table 1
Properties of ISO 4113 calibration oil.

| Density | Viscosity | Surface Tension |
|---------|-----------|-----------------|
| 821 kg/m$^3$ | 0.0032 kg/(m·s) | 0.02547 N/m |